\documentclass{pasj01}
\Received{$\langle$March 30, 2016$\rangle$}
\Accepted{$\langle$May 31, 2016$\rangle$}
\Published{$\langle$publication date$\rangle$}
\usepackage{color}

\begin{document}

\title{Rapid Merger of Binary Primordial Black Holes: an Implication for GW150914}
\author{Kimitake Hayasaki \altaffilmark{1,5}}%
\altaffiltext{1}{Department of Astronomy and Space Science, Chungbuk National University, Cheongju 361-763, Korea}
\email{kimi@cbnu.ac.kr}
\author{Keitaro Takahashi,\altaffilmark{2,5}}
\altaffiltext{2}{Department of Physics, Kumamoto University, Kumamoto 860-0862, Japan}
\author{Yuuiti Sendouda, \altaffilmark{3,5}}
\altaffiltext{3}{Graduate School of Science and Technology, Hirosaki University, Hirosaki 036-8561, Japan}
\author{Shigehiro Nagataki \altaffilmark{4,5}}
\altaffiltext{4}{Astrophysical Big Bang Laboratory, RIKEN, Saitama 351-0198, Japan}
\altaffiltext{5}{Yukawa Institute for Theoretical Physics, Kyoto University, Kyoto 606-8502, Japan}


\maketitle

\begin{abstract}
We propose a new scenario for the evolution of a binary of 
primordial black holes (PBHs). We consider a dynamical friction 
by ambient dark matter, scattering of dark matter particles with 
a highly eccentric orbit besides the standard two-body relaxation 
process to refill the loss cone, and interaction between the binary 
and a circumbinary disk, assuming that PBHs do not constitute the 
bulk of dark matter. Binary PBHs lose the energy and angular 
momentum by these processes, which could be sufficiently efficient 
for a typical configuration. Such a binary coalesces due to the 
gravitational wave emission in a time scale much shorter than the age
of the universe. We estimate the density parameter of the
resultant gravitational wave background. Astrophysical implication
concerning the formation of intermediate-mass to supermassive 
black holes is also discussed.
\end{abstract}

%
\section{Introduction}
%

Primordial black holes (PBHs) are hypothetical objects formed
due to gravitational collapse of density perturbations at
the early phase of the universe
\citep{Hawking:1971ei,Carr:1974nx}. A PBH mass larger than
$ \sim 10^{-18} M_{\odot} $, where $M_{\odot}$ is the solar mass,
is enough to prohibit evaporation via the Hawking radiation \citep{Hawking:1974rv}.
PBHs with such a mass could remain as the dark matter in the universe until now.

The composition of the dark matter has been poorly known although nowadays people 
are well aware that it amounts to more than $ 20\,\% $ of the whole energy in the universe 
{\citep{2011ApJS..192...16L,2015arXiv150201589P}}. The possibility that PBHs constitute 
a certain fraction of the whole dark matter has been investigated by many authors (see 
\cite{Carr:2009jm} and references therein). Some of those massive PBHs may be residing 
in Milky Way's halo as massive compact halo objects (MACHOs), but their abundance has 
been constrained for a certain range of mass by, e.g., gravitational lensing observations 
{\citep{Tisserand:2006zx,2014ApJ...786..158G}}.

\citet{Nakamura:1997sm} considered
the formation and evolution of binary PBHs. If PBHs were
formed randomly rather than uniformly in space, some pairs
of PBHs would have sufficiently small separations to overcome
the cosmological expansion and then form a bound system, binary.
They estimated the amplitude of gravitational waves from 
coalescing black holes following the binary evolution taking 
the gravitational wave emission into account. 
Because the gravitational radiation is not effective to extract an angular momentum from binary PBHs 
unless the separation is very small, only a small fraction of the binaries coalesces during the age of
the universe. Nonetheless, the cosmological gravitational wave background amounts to
$\Omega_{\rm gw} \sim 10^{-10}$ at the {\it Laser Interferometer Space Antenna} ({\it LISA}) band ($\nu_{\rm gw} \sim 10^{-4}~{\rm Hz}$), where $\Omega_{\rm gw}$
and $\nu_{\rm gw}$ are the density parameter and frequency
of the gravitational wave, respectively \citep{Ioka:1998gf, Inoue:2003di}

{
Recently, 
the Advanced Laser Interferometer Gravitational-Wave Observatory (LIGO) 
has detected the gravitational waves originated from the merger of nearly 
equal-mass binary black holes with $\sim30\,M_\odot$ at $z\approx0.09$
\citep{2016PhRvL.116f1102A}.
The discovery of this gravitational wave source (GW150914) has just brought 
to us a new problem how such an intermediate-mass binary is formed 
\citep{2016ApJ...818L..22A}. It remains a matter of debate, although some 
ideas have been already proposed. Binary PBHs could be one of candidates 
for merging massive binaries.
}

In this paper, we consider two prominent processes which affect the evolution 
of binary PBHs. One is the gravitational interaction between the binary and an 
ambient dark matter. Another is the gravitational interaction between the binary 
and a gaseous disk surrounding it (i.e.\ circumbinary disk). Both processes are 
inevitable and help the binary shrink rapidly. Our scenario predicts that binary 
PBHs generically coalesce during the age of the universe and gravitational waves 
are emitted more efficiently compared to the result in {\citet{Ioka:1998gf}}. Throughout 
this paper, we adopt the cosmological parameters obtained by 7-\rm{yr} {\it Wilkinson 
Microwave Anisotropy Probe} ({\it{WMAP}}) {\citep{2011ApJS..192...16L}}

In Section~\ref{section:binary-PBH}, we summarize the basic
elements of the evolution of binary PBHs, assuming that 
PBHs do not constitute the bulk of the dark matter. 
We argue the gravitational interaction of dark matter
and gas with the binary in Sections~\ref{section:DM}
and \ref{section:gas}, respectively. Finally, we summarize 
our results and also give a simple estimation of the gravitational 
wave background predicted in our scenario in 
Section~\ref{section:summary}.

%
\section{Binary PBHs \label{section:binary-PBH}}
%
First we give a summary on the evolution scenario of binary PBHs 
presented by \citet{Nakamura:1997sm} and \citet{Ioka:1998nz}, 
extending to a case where PBHs do not constitute the bulk of the 
dark matter.

Density fluctuations of radiation with a wave length exceeding the 
Hubble radius in the very early stage of the radiation-dominated 
era eventually come into the Hubble horizon and, if the amplitude 
is large enough (and not too large), gravitationally collapse to form 
PBHs. The black hole mass is comparable to the horizon mass and 
is written as $ M_{\rm{bh}} \sim 0.5\,\gamma\,(T_\mathrm{bhf}/\mathrm{GeV})^{-2}\,M_\odot $, 
where $T_\mathrm{bhf}$ is the temperature of the universe at the 
moment of the PBH formation and $\gamma$ is a numerical factor 
which, in an analytical treatment, takes a value around $ 0.2 $ \citep{Carr:1975qj}. 
Hereafter we suppress $ \gamma $ since we do not need very precise figures.
Also, \citet{1998PhRvD..57.6050K} showed the mass scales of PBHs is $\sim1\,M_\odot$ 
in a certain parameter region of our double inflation model. We, therefore, adopt the typical 
mass of PBHs as $1\,M_\odot$ in what follows.

For simplicity, we assume a primordial spectrum of density
fluctuations which has a sharp peak with a sufficient amplitude
to form PBHs so that they have a monochromatic mass function
(see, e.g., \cite{Kawaguchi:2007fz} for such an inflationary
model). 

We also assume that PBHs constitute a fraction, $f$, of the dark matter, that is, $\Omega_{\rm bh} = f \Omega_{\rm DM}$
where $\Omega_{\rm bh}$ and $\Omega_{\rm DM}$ are the current density parameter of PBHs and dark matter, respectively.
{Note that a certain fraction of the dark matter is in the form of particles with mass m much smaller than that of PBHs.}
The value of $f$ is constrained for a wide range of $M_{\rm bh}$ and can be as large as $0.1$ for
$M_{\rm bh} \lesssim 0.1 M_{\odot}$ while much severer
constraints, $f \lesssim 10^{-8}$, are obtained for
$M_{\rm bh} \gtrsim 10^3 M_{\odot}$ {\citep{Tisserand:2006zx,2008ApJ...680..829R, 
2013PhRvD..87b3507C, 2013PhRvD..87l3524C, 2014PhRvD..90j3522D, 2013PhRvD..88d1301P,
2014JCAP...06..026P,2014ApJ...786..158G}}. {Henceforth, we assume that $f\ll1$.}

Noting that the energy density of PBHs can be written as 
$\rho_{\rm bh}(T) = f \rho_{\rm DM}(T)=f\Omega_{\rm{DM}}(T/T_0)^3(3H_{0}^2/8\pi G)c^2$ 
where $T_{0}$, $H_0$, $c$ and $G$ are the temperature at the present universe, current 
Hubble parameter, light velocity, and gravitational constant, respectively, the average separation 
of PBHs at the formation, $\bar{r}$, is expressed as
\begin{equation}
\bar{r}
\sim \left( \frac{M_{\rm bh}c^2}{\rho_{\rm bh}(T_{\rm bhf})}
     \right)^{1/3}
\sim 8\times10^{-11} f^{-1/3}
     \left( \frac{M_{\rm bh}}{M_{\odot}} \right)^{5/6} {\rm pc},
\end{equation}
where {the normalization corresponds to $T_\mathrm{bhf}=1\,{\rm GeV}$.}
Actually, PBHs are randomly distributed in space at the formation
and we consider a pair of nearest PBHs with the initial separation
parametrized as $\alpha \bar{r}$. Statistically, most of nearest
PBH pairs will have $\alpha \lesssim 1$ while only a few would
have $\alpha \ll 1$. In fact, the probability density
distribution function is
$dP/d\alpha \sim 3 \alpha^2 e^{-\alpha^3}$
for a random distribution as given by \citet{Nakamura:1997sm}.

After the black hole formation, the separation of the pair evolves
in proportion to the scale factor due to the cosmic expansion.
However, if the local energy density of the pair,
$\rho_{\rm pair} = \alpha^{-3} \rho_{\rm bh}$, becomes larger
than the radiation energy density, $\rho_{\gamma}$, the pair
decouples from the cosmic expansion and forms a gravitationally
bound system. We see that the turnaround occurs at the temperature,
\begin{eqnarray} 
T_{\rm ta} \approx \left(\frac{\alpha^{3}}{f}\right)^{-1}T_{\rm eq}
\label{eq:zta}
\end{eqnarray}
{
with $T_{\rm eq}\sim1\,{\rm eV}$. 
}
Here it should be noted that for $\alpha > f^{1/3}$ we have
always $\rho_{\rm pair} < \rho_{\rm DM}$ so that the turnaround
never occurs. Thus we obtain the condition for $\alpha$ that 
a pair forms a bound system,
$\alpha_0 \leq \alpha \leq f^{1/3}$,
where
$\alpha_{0}
\sim 10^{-3} f^{1/3} (M_{\rm{bh}}/ M_{\odot})^{1/6}$
and the lower bound comes from the requirement that
$\alpha \bar{r}$ should be greater than the Schwarzschild
radius of a PBH. 
Hereafter we consider a typical pair with $\alpha \sim f^{1/3}$, 
which turns around at around the matter-radiation equality.
The separation at the turnaround is,
\begin{eqnarray}
r_{\rm ta}
&=& \frac{T_{\rm bhf}}{T_{\rm ta}} \alpha \bar{r}
\sim 5\times10^{-2}
       \left( \frac{\alpha^3}{f} \right)^{4/3}
       \left( \frac{M_{\rm bh}}{M_{\odot}} \right)^{1/3}
       {\rm pc}.
\label{eq:r_ta}
\end{eqnarray}

After the turnaround, if the two black holes had no relative
velocity, they would coalesce to form a single black hole.
However, as discussed in \citet{Nakamura:1997sm} and \citet{Ioka:1998nz},
the tidal force from neighboring black holes would give
the pair some angular momentum and prevent the coalescence. 
Since the semi-major axis of the binary is roughly given by $a \sim r_{\rm ta}$, 
the acceleration due to the tidal force and free fall time are estimated as,
\begin{equation}
a_{\rm tidal}
\sim \frac{G M_{\rm bh}}{R^2} \frac{r_{\rm ta}}{R}, ~~~
t_{\rm ff} \sim \sqrt{\frac{r_{\rm ta}^3}{G M_{\rm bh}}},
\label{tidal}
\end{equation}
respectively, where $R \equiv (T_{\rm bhf}/T_{\rm ta}) \bar{r}$
is the typical separation between the binary and the third PBH
at the turnaround of the pair. Thus the semi-minor axis
is estimated as
$b \sim a_{\rm tidal} t_{\rm ff}^2 \sim r_{\rm ta}^4/R^3\sim \alpha^3 r_{\rm ta}$.
We can see that the orbital eccentricity can be written as 
$e = \sqrt{1-b^2/a^2} \sim \sqrt{1-\alpha^6}$
(the typical value of $e$ is $\sqrt{1-f^2}$).

It is possible that the black hole pair obtains an angular
momentum from ambient density fluctuation of cosmic plasma
as well as the neighboring black holes. This effect can be
estimated as follows. Although the density fluctuations exist
for a wide range of length scale, the most effective would be
the one with scale of $r_{\rm ta}$. Denoting the amplitude
of fluctuation as $\delta \approx 10^{-5}$, we can think that
the black hole pair obtains angular momentum from an object
with mass $\rho_\gamma r_{\rm ta}^3 \delta$ which is separated
from the pair by $r_{\rm ta}$. Here $\rho_\gamma = \Omega_{\rm{\gamma}}(T/T_0)^4(3H_0^2/8\pi G)c^2$, 
where $\Omega_{\gamma}$ is the current density parameter 
of radiation, is the radiation energy density and the acceleration 
due to the tidal force from this object is written as,
\begin{equation}
a_{\rm tidal, fluc}
\sim G\rho_\gamma r_{\rm ta} \delta.
\end{equation}
Comparing this with the acceleration from a neighboring
black hole, equation (\ref{tidal}), we see that the effect of
the density fluctuations is important only when $f < \delta$
assuming $\alpha = f^{1/3}$.
It should be noted that peculiar velocities of PBHs are
negligible because density fluctuations already damped
at such a small scale as given by equation (\ref{eq:r_ta}).

Once a binary is formed, it emits gravitational waves
and the separation shrinks gradually. The coalescence
timescale of a binary PBH is given by \citet{p64},
\begin{eqnarray}
t_{\rm gw}
&=& \frac{5c^5a^4}{64 G^3 M_{\rm bh}^3} (1-e^2)^{7/2} 
\nonumber \\
&\sim&
     1\times 10^{34} f^7
     \left( \frac{\alpha^{3}}{f} \right)^{16/3}
     \left( \frac{M_{\rm bh}}{M_{\odot}} \right)^{-5/3}
     \left( \frac{a}{r_{\rm ta}} \right)^4
     {\rm yr},
\label{eq:t_gw}
\end{eqnarray}
{where we assumed that tidal force from a neighboring black hole
is dominant compared with that from density fluctuations,
$a_{\rm tidal} > a_{\rm tidal, fluc}$.
}
Thus we see that the gravitational wave emission is not
effective for most of the binaries 
to merge during the age of the universe, unless $f$ is
very small $f \lesssim 10^{-3.5} (M_{\rm bh}/ M_{\odot})^{5/21}$.
However, as we see below, considering two other processes which
extract the energy and angular momentum from the binary,
its evolution drastically changes. 
Note that if $f$ is very small, the number of the binaries
is very small so that the total amount of gravitational waves
is also small even if most of the binaries merge during
the age of the universe.

%
\section{Interaction with dark matter \label{section:DM}}
%

In this section, we shall describe how moving PBHs interact 
with the surrounding dark matter. In the case of $f \ll 1$, we 
consider here, the dark matter other than PBHs is abundant 
and will accrete onto PBHs forming a halo around them. 

At the turnaround of a pair of PBHs, each PBH will have a dark 
halo with the density $\sim \rho_{\rm DM}(T_{\rm ta})/c^2$.
It would be reasonable to assume that, at the turnaround, two 
dark halos around PBHs merge and collapse into a single halo 
around the binary. The halo would then undergo a violent relaxation 
and reach the virial equilibrium with the mass 
$M_{\rm DM}\sim\rho_{\rm DM}(T_{\rm{ta}})r_{\rm ta}^3/c^2\sim(\alpha^3/f)M_{\rm bh}$. 
Adopting a singular isothermal sphere solution, $\rho_{\rm{halo}}(r)=\sigma^2/{(2\pi Gr^2)}$, 
for the halo density profile, the velocity dispersion of dark halo and 
virial radius are $\sigma\sim0.2\times(\alpha^3/f)^{-1/6}(M_{\rm{bh}}
/M_{\odot})^{1/3}\rm{km}\,\,\mathrm s^{-1}$ and $r_{\rm{v}}=GM_{\rm{DM}}/\sigma^2=2r_{\rm{ta}}$, respectively. 

In the presence of enveloping dark matter, 
the separation of the two PBHs decays due to the dynamical friction, i.e., 
the energy of each PBH is lost by the gravitational interaction with the field 
dark matter particles. Its timescale is given by \citet{gd87} as
\begin{eqnarray}
t_{\rm df}
&\sim&\frac{v_{\rm bh}^3}
            {4 \pi G^2 M_{\rm bh} \rho_{\rm halo}(r)\ln{\Lambda}}
       \left[ {\rm erf}(X) - \frac{2X}{\sqrt\pi} e^{-X^2}       \right]^{-1}
\nonumber \\
&\sim&
  5 \times 10^3
  \left(\frac{\ln\Lambda}{\ln 10^6}\right)^{-1}\,
  \left(\frac{\alpha^3}{f}\right)\,\rm{yr},
\label{eq:tdf}
\end{eqnarray}
where $v_{\rm{bh}}=\sqrt{GM_{\rm{bh}}/r_{\rm{ta}}}\sim0.3(\alpha^3/f)^{-2/3}(M_{\rm{bh}}/M_\odot)^{1/3}$, 
${\rm erf}$ is the error function, and $X \equiv v_{\rm bh}/(\sqrt 2 \sigma)=1$.
The Coulomb logarithm, $\ln\Lambda$, is related, as 
$\ln\Lambda=\ln(r_{\rm{ta}}\sigma^2/2Gm)=\ln(N/2)$ where $m$ 
is the mass of the dark matter particle, to the number of dark matter particles {within the halo}:
\begin{equation}
N
= \frac{M_\mathrm{DM}}{m}
\sim
  10^6\,
  \left(\frac{\alpha^3}{f}\right)\,
  \left(\frac{m}{10^{-6}\,M_\odot}\right)^{-1}\,
  \left(\frac{M_{\rm{bh}}}{M_\odot}\right),
\label{eq:N}
\end{equation}
The binary orbit decays until the binding energy 
of the binary becomes comparable to the kinetic energy of the dark halo. 
Then, the binary becomes hard and the dynamical friction is no longer effective.
Equating the total kinetic energy of dark matter in the halo to the binding energy, 
we obtain the hardening radius, $a_{\rm h}=r_{\rm{ta}}/2$.

It should be here checked whether dynamical friction really works or not by 
comparing the timescale of dynamical friction, $ t_\mathrm{df} $, with the 
Hubble time $ H^{-1} $. In the matter-dominated era, we have (for $ \alpha \sim f^{1/3} $)
\begin{equation}
H\,t_\mathrm{df}
\sim
  \frac{1}{\Gamma}
  \left(\frac{1+z}{1+z_\mathrm{eq}}\right)^{3/2}\,,
\end{equation}
where $\Gamma = 1/[H(z_\mathrm{eq})\,t_\mathrm{df}] = \mathcal{O}(1-10)$ is a numerical factor.
Thus the dynamical friction could be in fact effective for $ 1+z < \Gamma^{2/3}\,(1+z_\mathrm{eq}) $, 
i.e., throughout the matter-dominated era.

After the binary becomes hardened, dark matter particles 
in the loss cone are depleted {(e.g. \cite{Merritt:2004gc})}.
The binary orbit can still, however, remain to decay due to 
the dark matter scattering if dark matter particles refill the 
loss cone by two-body relaxation. Its relaxation timescale is 
given by \citet{spitzer87} as
\begin{eqnarray}
t_{\rm{tbr}}(r)
&\approx&
\frac{0.34\sigma^3N}{G^2\rho_{\rm{halo}}(r)M_{\rm{DM}}\ln\Lambda}
\nonumber \\
&\sim&
5\times10^9
\left(\frac{r}{r_{\rm{v}}}\right)^{2}
\left(\frac{\alpha^3}{f}\right)^{3/2}
  \left(\frac{N/\ln N}{10^6/\ln 10^6}\right)
{\rm{yr}},
\label{tbr}
\end{eqnarray}
This is marginally shorter than the age of the universe, $ t_0 \approx 1.37 \times 10^{10}\,\mathrm{yr} $, for $ N $ of order $ 10^6 $, so there is a possibility for some range of $ N $ that the dark matter scattering continues to work to reduce the semi-major axis of the binary. 
Its orbital decay rate obeys
\begin{eqnarray}
\frac{d}{dt}\left(\frac{1}{a}\right)=-C\frac{G\rho_\mathrm{halo}}{\sigma},
\label{eq:hr}
\end{eqnarray}
where $C$ is a numerical factor, the typical value of which is 14.3 \citep{1996NewA....1...35Q}.
Integrating equation (\ref{eq:hr}) under the assumption of a constant-density core, we obtain
\begin{eqnarray}
a_{\rm{ds}}(t)
&=&\frac{\sigma}{CG\rho_{\rm{halo}}(a_{\rm{h}})t}
\sim2\times10^{-7}
\nonumber \\
&\times&
\left(\frac{\alpha^3}{f}\right)^{-8/3}
\left(\frac{M_{\rm{bh}}}{M_\odot}\right)^{1/3}
\left(\frac{t_{\rm{tbr}}(r_{\rm{v}})}{t}\right)
~\rm{pc}.
\end{eqnarray}
Unfortunately, the above scenario will not be realized in most cases as we shall explain.
Since $ t_\mathrm{tbr} $ is proportional to $N$ and $N$ increases as $m$ falls below {$10^{-6}\,M_\odot$} as seen in equation (\ref{eq:N}), 
it exceeds the age of the universe for lower masses of the halo dark matter.
In particular, if the bulk of dark matter is composed of elementary particles like supersymmetric ones, then binary PBHs are stalled at the hardening radius.

Here we advocate an alternative approach to make the binary orbit decay by supplying the dark matter particles into 
the loss cone radius. In our cosmological setup, the ambient dark matter continues to accrete onto the binary as its influence radius expands. A binary in the dark halo could in principle shrink by scattering dark matter particles. However, to do so, it is necessary that the dark matter particles pass close enough to the binary, that is, they enter the loss cone 
{\citep{Merritt:2004gc}}. How much the binary shrinks is determined by the dark matter mass supplied into the loss cone.

For simplicity, we assume that a dark matter particle enters the loss cone 
if the pericenter distance of its orbit is shorter than the semi-major axis of 
the binary. Let us evaluate the orbital elements of a typical dark matter particle.
As in the case of a PBH pair, we suppose that dark matter particles feel a tidal 
force from nearby PBHs during their infall. Since the turnaround radius of dark 
matter is given as $ r_{\mathrm{ta},\mathrm{dm}} \sim 10^{-2}\,(M_\mathrm{bh}/M_\odot)^{1/3}\,((1+z)/(1+z_\mathrm{eq}))^{-4/3}\,\mathrm{pc}$ {\citep{1985ApJS...58...39B,Mack:2006gz,2008ApJ...680..829R}}, the semi-major and semi-minor axes are estimated as $ \sim r_{\mathrm{ta},\mathrm{dm}} $ 
and $ \sim r_{\mathrm{ta},\mathrm{dm}}^4/R^3 $, respectively, where $ R $ is the 
distance to the third PBH at the moment of infall. Since the dark matter particles 
come to have highly eccentric orbits, we evaluate the pericenter distance as 
\begin{equation}
d
\sim
  \frac{r_{\mathrm{ta},\mathrm{dm}}^7}{R^6}
\sim
  10^{-2}\,f^2\,
  \left(\frac{1+z}{1+z_\mathrm{eq}}\right)^{-10/3}\,
  \left(\frac{M_\mathrm{bh}}{M_\odot}\right)^{1/3}\,
  \mathrm{pc}\,.
\end{equation}

Requiring that $ d < a \sim r_\mathrm{ta} $ yields the following condition on the redshift:
\begin{equation}
1+z
> (1+z_\mathrm{eq})\,
  \left(\frac{f}{\sqrt{5}}\right)^{3/5}\,
  \left(\frac{\alpha^3}{f}\right)^{-2/5}\,.
\end{equation}
Dark matter particles which turnaround before this redshift enters 
the loss cone of the binary in a dynamical time. Since the total mass 
of the whole dark matter which turns around by a redshift $z$ is 
given by $ \sim M_\mathrm{bh}\,(1+z_\mathrm{eq})/(1+z) $ 
{\citep{Mack:2006gz}}, the mass which refills the loss cone is 
estimated as
\begin{equation}
M_\mathrm{lc}
\sim
  \left(\frac{f}{\sqrt{5}}\right)^{-3/5}\,
  \left(\frac{\alpha^3}{f}\right)^{2/5}\,
  M_\mathrm{bh}\,.
\label{eq:M_lc}
\end{equation}

It should be kept in mind that the above calculation is crude and there were many uncertainties which could cumulatively affect the final expression of $ M_\mathrm{lc} $.
A precise evaluation will need some N-body simulations.
However, we stress that the primary implications of equation~(\ref{eq:M_lc}) that $ M_\mathrm{lc} $ is proportional to the black hole mass $ M_\mathrm{bh} $ and that it increases as the fraction $ f $ decreases seem robust.
We hence believe that this mechanism certainly has a potential to supply non-negligible amount of dark matter into the loss cone region around binary PBHs.

Now we can evaluate how much the binary could shrink due to the dynamical friction.
We parametrize the uncertainties in the above computation by a function $ \eta $ of 
the parameters $ f $ and $ \alpha $: $ M_\mathrm{lc} = \eta(f,\alpha)\,M_\mathrm{bh}$.
A dark matter element of mass $dm$ would extract an energy $ \sim G\,M_\mathrm{bh}\,dm/a $ 
from a binary {\citep{Merritt:2004gc}}, so the evolution of the orbital radius obeys
\begin{equation}
G\,M_\mathrm{bh}^2\,d\left(\frac{1}{a}\right)
=
  \frac{G\,M_\mathrm{bh}}{a}\,dm\,.
\end{equation}
Thus, the residual separation after the dynamical friction ceases is estimated as
\begin{equation}
a_\mathrm{df}
\sim
  \exp\left(-\frac{M_\mathrm{lc}}{M_\mathrm{bh}}\right)\,r_\mathrm{ta}
= e^{-\eta(f,\alpha)}\,r_\mathrm{ta}\,.
\end{equation}
If the function $ \eta $ acquires a value around $ 7 $, which, if equation (\ref{eq:M_lc}) 
applies, requires only a reasonable value $ f \approx 0.1 $, then the binary could shrink 
by about three orders of magnitude within $t_\mathrm{df}$.

Again we note that the shrinkage factor contains various uncertainties that come from 
the uncertainties in the determination of $ M_\mathrm{lc} $. Here we have restricted 
ourselves to proving the potential utility of our scenario, but will revisit this issue in more 
realistic setups via both analytical and numerical approaches in the future publications.

%
\section{Interaction with gas \label{section:gas}}
%

Baryon gas, as well as the dark matter, also accretes onto
the PBHs. Even if the gas has negligible angular momentum with
respect to the center of mass of the binary, the gas can not
fall directly into the black holes but instead would form
a rotating disk around the binary (i.e. a circumbinary disk).
Many authors have been addressed the computational 
hydrodynamic simulations of interaction between the nearly 
equal-mass binary black holes and the circumbinary disk 
\citep{2007PASJ...59..427H,2008ApJ...682.1134H,2009MNRAS.393.1423C,roedig11,shi12,fa14}. 
The circumbinary disk can play a role of an efficient mechanism
to extract the angular momentum from a binary
\citep{1991ApJ...370L..35A,an05,mm08,Hayasaki:2008eu,2009MNRAS.393.1423C,do13}.
This is mainly because the circumbinary disk and nearly equal-mass binary exchange their
masses and angular momenta through the tidal-resonant interaction
\citep{1994ApJ...421..651A}.

Neglecting the angular momentum of the gas, we consider
the Bondi accretion and its accretion radius is given by
\begin{equation}
r_{\rm B}
= \frac{G M_{\rm bh}}{c_{\infty}^2}
\sim 4 \times 10^{-5}
     \frac{M_{\rm bh}}{M_{\odot}}
     \frac{1+z_{\rm eq}}{1+z}~\rm{pc},
\label{r_B}
\end{equation}
where $c_{\infty} \sim 10 \sqrt{(1+z)/(1+z_{\rm{eq}})}~\rm{km/s}$
is the average sound velocity {\citep{Ricotti:2007jk}}.
When the Bondi radius is smaller than the separation of
the binary, the gas accretes onto each PBH separately
and a circumbinary disk would not form. 
Because the Bondi radius increases in time while the binary separation would
shrink due to the dynamical friction, the Bondi radius
exceeds the inner edge of circumbinary disk, 
which is typically given by twice the binary separation \citep{1994ApJ...421..651A}, 
sooner or later and a circumbinary disk would form.
The circumbinary disk would begin to form at $z \lesssim 400$ when $r_{\rm{B}}=a_{\rm{ds}}$ or $r_{\rm{B}}=a_{\rm{df}}$.

According to \citet{Ricotti:2007jk} and \citet{2008ApJ...680..829R},
the gas accretion rate,
{$\dot{M}_{\rm B}$,}
onto a black hole with $1 M_{\odot}$ has a peak
around the matter-radiation equality time and the peak accretion
rate is several percent of the Eddington rate,
$\dot{M}_{\rm Edd} \sim
 2 \times 10^{-9} (M_{\rm bh}/M_{\odot}) M_{\odot} {\rm /yr}$.
For $M_{\rm bh} \leq M_{\odot}$, the mass of the circumbinary disk,
{$M_{\rm cbd} \sim \dot{M}_{\rm B} H^{-1}$},
can be well fitted
by $M_{\rm cbd}/M_{\rm bh} \sim 10^{-5} (M_{\rm bh}/M_{\odot})$.
Considering only the interaction between the binary and the
circumbinary disk, that is, neglecting the effects of the respective
disks, the time scale of the orbital decay can be estimated by 
\citet{Hayasaki:2008eu} (see also equations (16)-(19) of \cite{2010PASJ...62.1351H} 
and alternatively equation (38) of \cite{2013PhRvD..87d4051H}) as
\begin{eqnarray}
t_{\rm cbd}
&=& t_{\rm vis} \frac{M_{\rm bh}}{M_{\rm cbd}}
\nonumber \\
&\sim& 7 \times 10^6
       \left( \frac{a}{10^{-4}~{\rm pc}} \right)^{1/2}
       \left( \frac{M_{\rm bh}}{M_{\odot}} \right)^{-1}
       \rm{yr},
\label{eq:t_gas}
\end{eqnarray}
where
$t_{\rm vis} \sim 7 \times 10^3 (a/{\rm pc})^{1/2} \rm{yr}$
is the viscous time scale of the circumbinary disk with
an assumption that the temperature of the inner edge of
the circumbinary disk is $(1+z) T_{0}$ and the Shakura-Sunyaev
viscosity parameter is $0.1$ \citep{1973A&A....24..337S}.

As the separation decreases, the time scale given by equation (\ref{eq:t_gas})
also decreases gradually while the timescale of gravitational waves,
equation (\ref{eq:t_gw}), decreases much more rapidly. When the two time
scales become comparable, the most effective process of extracting
angular momentum from the binary switches to the gravitational wave
emission. The critical separation, $a_{\rm c}$, is,
\begin{equation}
a_{\rm c}
\sim 10^{-9}
     \left( \frac{M_{\rm bh}}{M_{\odot}} \right)^{4/21}
     \rm{pc},
\label{eq:a_c}
\end{equation}
and the critical redshift, $z_{\rm{c}}$, which is determined
by $H^{-1} = t_{\rm cbd}$ is given by,
\begin{equation}
1+z_{\rm c}
\sim 100 \left( \frac{M_{\rm bh}}{M_{\odot}} \right)^{2/3}.
\end{equation}
Here we assumed the binary orbit is circular after the dynamical
friction. Substituting equation (\ref{eq:a_c}) into equation (\ref{eq:t_gw}),
the time scale of gravitational radiation reduces to
$t_{\rm gw} \sim 10^4 (M_{\rm bh}/ M_{\odot})^{-19/21}{\rm yr}$.
Thus, for the typical parameter set, PBH binaries of
$M_{\rm bh} \gtrsim 0.01 M_{\odot}$ generically coalesce
by around $z_{\rm c} > 5$. This is to be contrasted
with \citet{Nakamura:1997sm} where only a small fraction
of binary PBHs coalesces during the age of the Universe.

%
\section{Summary and Discussions \label{section:summary}}
%

We have studied the evolution of binary PBHs by considering
the gravitational interactions with the ambient dark matter and 
baryon gas. Let us summarize our scenario in turn. In the early 
stage of radiation dominated era, the PBHs are formed by the 
gravitational collapse of radiation with the prominent amplitude 
of the density fluctuations. Some PBH pairs of randomly distributed 
PBHs decouple from the cosmic expansion and form binary PBHs 
at the turnaround ($z \sim z_{\rm eq}$).

Each dark halo, which is formed by the dark matter accretion onto 
each PBH, collapses into a single halo around binary PBHs. The 
energy and angular momentum of the binary are lost by the dynamical 
friction against the dark matter particles in the halo. The dynamical 
friction is not so effective when the binary is hardened at the half 
radius of the turnaround radius. Even after the hardening, the scattering 
between binary PBHs and infalling dark matter particles by two-body 
relaxation becomes effective. The binary orbit then decays by about five 
orders of magnitude of the turnaround radius within the age of universe.
This mechanism, however, seems not to work in the realistic case because 
of the strong constraint on the mass of dark matter particle (should be 
$m\lesssim10^{-6}M_\odot$) by the recent {EROS-2 \citep{Tisserand:2006zx} 
and Kepler \citep{2014ApJ...786..158G} observations}.

If the third PBH {tidally affects} the orbits of dark matter particle around 
the binary, their orbits could become highly eccentric. A family of dark 
matter particles with high eccentric orbits makes the binary shrink effectively 
and rapidly. Although a precise estimation of how much shrinking the binary 
separation is hard analytically, our simple model for $f\approx0.1$ implies 
that the binary orbit can decay by three orders of magnitude of the turnaround 
radius.

Some time after the recombination ($z \lesssim 1000$),
the circumbinary disk would form and extract the angular momentum
from the binary. At $z \sim O(100)$, the most effective process
extracting the angular momentum from the binary switches to
the gravitational wave emission. Finally the binary would
coalesce at around $z \sim 100$.

If our scenario really works, most of binary PBHs coalesce around
$z = z_{\rm c}$ emitting gravitational waves. Thus we can expect
a substantial amount of the gravitational wave background. 
Below we present an order-of-magnitude estimation
simply assuming that all binaries coalesce at $z = z_{\rm c}$.
Given the probability distribution function for the initial
separation, $dP/d\alpha \sim 3 \alpha^2 e^{-\alpha^3}$,
we see that the fraction of PBHs which form a binary is 
$P(\alpha < f^{1/3}) \sim f$ assuming $f \ll 1$.
Then, at $z = z_{\rm c}$, the number of binaries in a horizon
volume is given by $\sim f^2 H^{-3} \rho_{\rm DM} / M_{\rm bh}$.
Noting that the energy of gravitational waves emitted
by a binary is $\sim (G M_{\rm bh} \nu)^{2/3} M_{\rm bh}$,
where $\nu = \sqrt{G M_{\rm bh} / a^3}$ is the frequency of
the gravitational wave determined by the binary separation,
the energy density of the gravitational waves at
$z = z_{\rm c}$ is
$\epsilon_{\rm gw}(\nu,z_{\rm c})
 \sim (G M_{\rm bh} \nu)^{2/3} f^2 \rho_{\rm DM}(z_{\rm c})$.
Finally, the current density parameter of the gravitational
waves is given by,
\begin{eqnarray}
\Omega_{\rm gw,0}(\nu)
&=& \frac{\epsilon_{\rm gw}((1+z_{\rm c}) \nu,z_{\rm c})}
         {(1+z_{\rm c})^4 \rho_{\rm c}}
\nonumber \\
&\sim& 4 \times 10^{-7} f^2
       \left( \frac{\nu}{\nu_{\rm c,0}} \right)^{2/3}
       \left( \frac{M_{\rm bh}}{M_{\odot}} \right)^{37/63},
\end{eqnarray}
where
$\nu_{\rm c,0} \sim
  5 \times 10^{-2} (M_{\rm bh}/M_{\odot})^{3/14}
  (a/a_{\rm c})^{-3/2} / (1+z_{\rm c})\,{\rm Hz}$ represents
the lowest frequency corresponding to the separation
$a_{\rm c}$. The frequency domain is given as
$\nu_{\rm c,0} < \nu < \nu_{\rm mso, 0}$ where
$\nu_{\rm mso,0} \sim
 10^5 (M_{\rm bh}/M_{\odot})^{-1} / (1+z_{\rm c})\,{\rm Hz}$, 
which is the frequency corresponding to the marginally stable
orbit of a PBH binary. The gravitational wave background
with $f < 0.03$ is estimated as $\Omega_{\rm gw,0} \sim 10^{-10}$, 
which is comparable with that estimation by \citet{Nakamura:1997sm}
corresponding to $f=1$, and is enough large to be detected
with {{\it eLISA} \citep{as12}, {\it DECi-hertz Interferometer Gravitational wave 
Observatory} ({\it DECIGO}) \citep{seto01} and the Advanced LIGO \citep{2016PhRvL.116f1102A}}. Thus the observation 
of the gravitational wave background would give a strong
constraint on the dark matter fraction of PBHs with
$M_{\rm bh} \lesssim M_{\odot}$.

Let us mention the other astrophysical implications of our
scenario. Fermi detected gamma-rays with luminosity 
$\sim10^{49}\,\rm{erg}\,\rm{s^{-1}}$, $0.4\,\rm{s}$ after 
the black hole coalescence \citep{conn16}. 
The probability of chance coincidence is $0.002$ or so. 
\citet{loeb16} suggested that this electromagnetic 
counterpart might be triggerd by the collapse of a single, rapidly 
rotating massive star surrounding the binary black hole.
Alternatively it would suggest the existence of a circumbinary 
disk not decoupled from the binary intermediate-mass black holes 
(e.g., \cite{2013PhRvD..87d4051H}). If so, the short gamma-ray 
event is caused by the subsequent accretion of 
the circumbinary disk after the coalescence.

GW150914 has finally $62^{+4}_{-4}M_\odot$ after the gravitational wave 
burst by the merger of two black holes with $36^{+5}_{-4}$ and $29^{+4}_{-4}$ 
solar masses \citep{2016PhRvL.116f1102A}. Such massive black holes are 
thought to be formed by the metal poor star. 
Successive mergers of PBHs discussed in this paper could also explain 
the existence of the binary black holes with such intermediate masses.
Further studies of the formation channel of intermediate-mass black holes are desired
\footnote{
 \citet{sm16} have very recently suggested that the binary black holes having lead to GW150914 
 are a primordial origin by estimating the merger rate of a binary composing of two $30\,M_\odot$ 
 PBHs. They have not, however, taken into account the interaction processes with 
 surrounding dark matter and gasses which we proposed in this paper. We will come back to the 
 issue of how the binary of $30M_\odot$ black holes could form after successive mergers of 
 $\gtrsim1\,M_\odot$ PBHs within our scenario in a forthcoming paper.}.

If the spectrum of density perturbation is broad
rather than monochromatic as assumed here, clusters of PBHs
can be formed \citep{2006PhRvD..73h3504C}. If this is the case,
PBH mergers may occur frequently in clusters to form much larger
black holes than the original PBHs {\citep{CG2016,2016arXiv160300464B}}. 
This may help to explain the presence of a supermassive black hole recently discovered
at $z \sim 6$ \citep{Fan:2003wd,Willott:2003xf}, providing 
supermassive black holes directly by successive merger of 
PBHs or smaller ``seed'' black holes {\citep{2012PhLB..711....1K}}.

%
\section*{Acknowledgments}
%
KH is grateful to Shin Mineshige, Ryoji Kawabata, Kohta Murase,
Kiki Vierdayanti, and Junichi Aoi for helpful discussions.
This work has been supported in part by the Ministry of Education,
Science, Culture, and Sport and Technology (MEXT) through
Grant-in-Aid for Scientific Research (19740100, 18104003, 21540304, 
22540243, 22340045 [KH], 24340048, 26610048, 15H05896 [KT], 
26800115, JP16K17675 [YS], 19104006, 19740139, 19047004, 25610056, 26105521, 
26287056 [SN]). KH was supported in part by the research grants of the 
Chungbuk National University in 2015 and Korea Astronomy 
Space Science Institute in 2016.
KT was supported in part by Monbukagaku-sho
Grant-in-Aid for the global COE programs, ``Quest for
Fundamental Principles in the Universe: from Particles to
the Solar System and the Cosmos'' at Nagoya University.
YS was supported in part by JSPS through Grant-in-Aid for JSPS Fellows.

\end{document}